\documentclass[aps,prd,twocolumn,amsmath,amssymb,nofootinbib,longbibliography,nofootinbib]{revtex4-1}
\usepackage{amsfonts,amsthm,mathrsfs}
\usepackage{lipsum,graphicx}
\usepackage[usenames,dvipsnames]{color}
\usepackage{inputenc}
\usepackage{bm}
\usepackage[normalem]{ulem}
\usepackage{multirow}
\usepackage{float}
\usepackage{url}
\usepackage{natbib}
\usepackage{subfigure}
\usepackage[colorlinks=true,citecolor=Cerulean,urlcolor=Cerulean,linkcolor=Cerulean]{hyperref}
\begin{document}
\preprint{}
\title{Bose-Einstein condensate stars in combined Rastall-Rainbow gravity}
\author{O. P. Jyothilakshmi}
\email{op\_jyothilakshmi@cb.students.amrita.edu}

\author{Lakshmi J. Naik}
\email{jn\_lakshmi@cb.students.amrita.edu}

\author{V. Sreekanth}
\email{v\_sreekanth@cb.amrita.edu}

\affiliation{Department of Sciences, Amrita School of Physical Sciences, Amrita Vishwa Vidyapeetham, Coimbatore 641112, India}

\date{\today}
\begin{abstract}
We study  
zero and finite temperature static Bose-Einstein condensate (BEC) stars in the combined Rastall-Rainbow (RR) theory of gravity 
by considering different BEC equation of states (EoSs). 
We obtain the global properties of BEC stars 
by solving the modified Tolman-Oppenheimer-Volkoff equations 
with values of Rastall parameter $\kappa$ and Rainbow function $\Sigma$ chosen accordingly to get the results in theories of Rastall, Rainbow and RR. 
We observe that the parameter $\kappa$ has negligible effect on the maximum mass 
of the stars considered, whereas $\Sigma$ alters it significantly,  
and increasing the value of $\kappa$ beyond a certain limit results in unstable solutions for any value of $\Sigma$. We report that the inclusion of temperature in our analysis expands the parameter space by including more values of $\kappa$. However, temperature has negligible effect on the maximum mass of the stellar profiles in all the three theories. 
We find that the maximum masses and radii of the stars within RR theory can have good agreement with the observational data on pulsars for all the EoSs considered and in particular, the Colpi-Wasserman-Shapiro 
EoS, which was ruled out in General Relativity (GR). 
We also find that, 
in contrast to the results of GR, BEC stars consistent with observations can be realised in the RR theory with smaller bosonic self-interaction strength.
\end{abstract}
\maketitle

The theory of General Relativity (GR), proposed by Albert Einstein more than a century ago~\cite{Einstein:1916vd}, is the current standard theory of gravity which relates the 
curvature of the space-time to the energy and momentum of matter present in it.
The theory has successfully explained many aspects of the known universe. For example, the detection of gravitational waves by binary black holes~\cite{BHbinary}, neutron stars~\cite{NSbinary} and the recent observation of first shadow of a black hole obtained by the Event Horizon Telescope~\cite{EventHorizon2019}. 
However, GR has limitations in certain astrophysical and cosmological scenarios such as dark matter, dark energy and early inflation. These limitations of GR are some of the reasons for the proposal of various modified theories of gravity~\cite{Clifton:2011,Nojiri:2017ncd}. 
Out of such theories, the Rastall~\cite{Rastall72} and Rainbow~\cite{Magueijo2004} and combined Rastall-Rainbow~\cite{Mota:2019} formalisms have attracted a lot of interest in the fields of astrophysics and cosmology in recent times~\cite{Batista2011,Bronnikov:2016,Heydarzade:2016,Nashed2023,Magueijo2004,Galan:2006,Hackett:2005mb,Junior:2020zdt,Mota:2019}. 
\par 
Rastall's theory of gravity  can be stated as a generalisation of GR with the conservation law of energy-momentum tensor in curved space-time modified as \cite{Rastall72}
\begin{eqnarray}
{T^{\mu\nu}}_{;\mu} = \frac{1-\kappa}{16\pi G}R^{,\nu}, 
\label{eq:Rconservacao}
\end{eqnarray}
where $R$ is the Ricci scalar and $G$ is the Newtonian gravitational constant (throughout this paper, we keep the velocity of light $c=1$). Here, the Rastall parameter $\kappa$ is a constant and it measures the deviation of the theory from GR. 
If we set $\kappa =1$, the field equations of GR are obtained and $R=0$ gives the conservation law for flat space-time. 
It must be noted that, there have been claims on the 
equivalence of Rastall gravity and Einstein's GR~\cite{Visser:2017gpz,Golovnev:2023evp} and studies refuting it~\cite{Darabi:2017coc}. 
Gravity's rainbow~\cite{Magueijo2004} is proposed as an extension of doubly special relativity~\cite{Magueijo:2002} to curved space-time. Within the theory, the energy $(E)$-momentum $(p)$ relation of a test particle of mass $m$ is written as~\cite{Magueijo2004}
\begin{equation}
    E^{2}\Xi(x)^{2}-p^{2}\Sigma(x)^{2}=m^{2},
\label{eq1}
\end{equation}
where $x=E/E_{p}$ is the ratio of energy of the test particle to the Planck energy $E_{p} = \sqrt{\hbar /G}$ with $\hbar$ being the reduced Planck's constant. 
Here, $\Xi(x)$ and $\Sigma(x)$ are arbitrary energy dependent functions called the Rainbow functions. Under a low energy regime ($x\rightarrow 0$), the rainbow functions $\Xi (x)$ and $\Sigma (x)$ approach to unity, 
which recovers the standard dispersion relation. 
In this theory, the space-time depends on the energy of the particle, which gives different geometries in the same inertial frame for particles with different energy. 
The theory could fix the horizon problem since the cosmological solutions result in an energy dependent age of the universe~\cite{Magueijo2004}. 
\par 
Rastall-Rainbow (RR) theory, as the name suggests, is another modified theory in which the effects of both the Rastall and Rainbow gravities are combined together into a single formalism~\cite{Mota:2019}. In this alternate theory, the Einstein field equations are modified 
by considering an energy dependent metric and gravitational constant; 
\begin{equation}
  R_{\mu\nu}(x)-\frac{\kappa}{2}g_{\mu\nu}(x)R(x)= 8\pi G (x)T_{\mu\nu}(x)    \label{eq10}. 
\end{equation}
Within RR theory, spherically symmetric static solutions were obtained and used to study neutron stars~\cite{Mota:2019}. 
\par 
We now turn our attention to the formation of stellar equilibrium in these modified  theories. For a static system, in the case of GR, the stellar equilibrium conditions are 
described by the well known Tolman-Oppenheimer-Volkoff (TOV) equations~\cite{Tolman-1939,Oppenheimer39}. These coupled differential equations are extensively used in the studies of compact objects. In Rastall gravity, the condition for static stellar equilibrium has been derived recently~\cite{Oliveira:2015lka}. 
Further, in this study they have considered neutron stars within this formalism. Properties of strange stars in Rastall's gravity have also been investigated in Ref.~\cite{Shahzad2019}. Apart from the static case, there has been attempts to study rapidly rotating 
compact stars~\cite{daSilva:2020} and tidal Love numbers~\cite{Meng:2021} within this modified gravity. Coming to the Rainbow formalism, the modified TOV equations were obtained and have been employed to study compact stars~\cite{Hendi_2016,Garattini:2016}. 
Recently, dark energy stars have been studied within Rainbow theory~\cite{Tudeshki:2022wed}.  
Further, the authors of Ref.~\cite{Mota:2019}, by studying the static configurations in combined Rainbow-Rastall gravity, obtained the modified version of the TOV equations and used it to analyse the global properties of neutron stars. Using this formalism, properties of gravastars~\cite{debnath2021} as well as charged anisotropic compact stars~\cite{Mota:2022} have been studied recently. As we can see different compact stars and many exotic stars have been investigated under these modified theories.
We aim to study the 
Bose-Einstein condensate (BEC)
stars within the RR gravity. 
\par 
In order to study the BEC stellar structure, we make use of the modified TOV equations derived in the context of Rastall-Rainbow theory~\cite{Mota:2019}. 
It is made possible by considering the energy dependent static spherically symmetric metric in $(t,r,\theta,\phi)$ coordinates \cite{Mota:2019}
\begin{equation}
g_{\mu\nu} =
\left[
\begin{matrix}
-B(r)/\Xi^{2} & 0 & 0 & 0\\
0 & A(r)/\Sigma^{2} & 0 & 0 \\
0 & 0 & r^{2}/\Sigma^{2} & 0 \\
0 & 0 & 0 & r^{2} \sin{\theta}^{2}/\Sigma^{2}
\end{matrix}
\right], \label{eq5}
\end{equation}
where $A(r)$ and $B(r)$ denote the radial functions.
Now, by considering the stellar matter as static perfect fluid with four-velocity $u_{\mu} = (\Xi(x)/\sqrt{B(r)},0,0,0)$, the RR field equation (Eq.~\eqref{eq10}) together with the above metric yield the modified TOV equations~\cite{Mota:2019}
\begin{eqnarray}
\label{eq:TOVPR}
\frac{d\tilde{p}}{dr} &=& -\frac{G \tilde{M} \tilde{\rho}}{r^2}\left[\frac{\left(1+\tilde{p}/\tilde{\rho}\right)\left(1+4\pi r^3 \tilde{p}/\tilde{M}\right)}{1 - 2G\tilde{M}/r}\right], \\
\frac{d\tilde{M}}{dr}&=&4\pi \tilde{\rho}\, r^2. 
\label{eq:TOVMR}
\end{eqnarray}
Here, the effective pressure $\tilde{p}$ and energy density $\tilde{\rho}$ are defined in terms of the Rastall parameter $\kappa$ and the Rainbow function $\Sigma$ as 
\begin{eqnarray}
    \tilde{\rho} &=& \frac{1}{\Sigma(x)^{2}}\left[\alpha_{1}\rho+3\alpha_{2}p\right],\label{rhotilde}\\
    \tilde{p} &=& \frac{1}{\Sigma(x)^{2}}\left[\alpha_{2}\rho+(1-3\alpha_{2})p\right]; \label{ptilde}
\end{eqnarray}
with, $\alpha_{1}=\frac{1-3\kappa}{2(1-2\kappa)},\,\alpha_{2}=\frac{1-\kappa}{2(1-2\kappa)}.$
The two free parameters of the theory, $\Sigma$ and $\kappa$ modify the mass of the star $\tilde{M}$, in this unified formalism. 
It must be emphasized that, by convenient choice of the values of $\kappa$ and $\Sigma$, stellar equilibria in Rastall $(\Sigma=1)$, Rainbow $(\kappa=1)$ and combined RR gravities can also be studied using the above modified TOV equations. Further, when we set both $\kappa=\Sigma=1$, the TOV equations of GR are retrieved.
We note that, the parameter $\kappa$ cannot have values in the range $1/2<\kappa<2/3$, since the stellar mass goes to negative within that range~\cite{Mota:2019}.
The stellar structure equations obtained in this formalism are similar to the TOV equations of GR. These equations are solved from the center ($r=0$) to the surface ($r=R$) of the star imposing the boundary conditions: $\tilde{M}(0) = 0$ and $\tilde{p}=p_c$. The pressure $\tilde{p}$ vanishes near the surface of the star, which gives the mass $\tilde{M}$ and corresponding radius $R$ of the star. We solve the coupled differential equations for different central pressures and obtain all possible stellar configurations for the model chosen i.e., BEC star in the present analysis. 
\par 
\begin{figure*}[ht]
  \centering
  \subfigure[]{\includegraphics[width=0.47\linewidth]{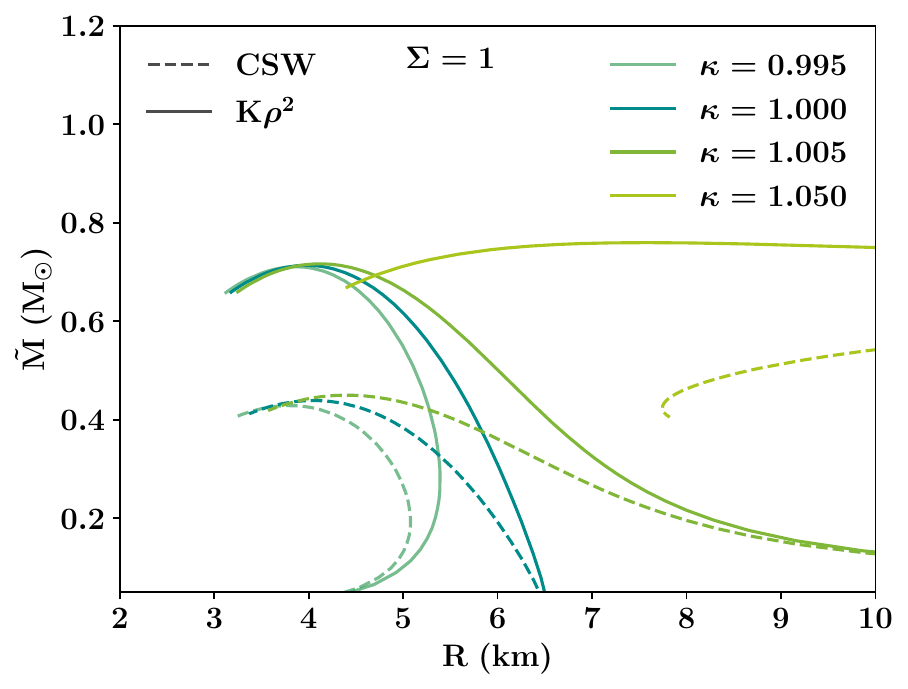}
  \label{MR-Rastall}} \quad
  \subfigure[]{\includegraphics[width=0.47\linewidth]{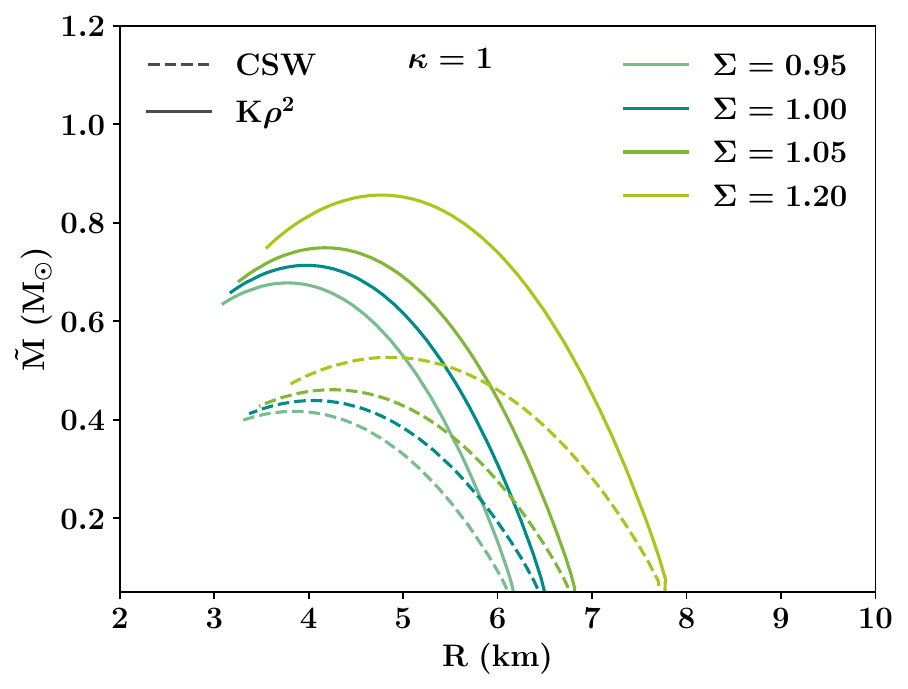} \label{MR-Rainbow}}
  \caption{
 Stellar configurations at $T=0$\,K for the two BEC EoSs considered within the (a) Rastall and (b) Rainbow gravities.}\label{Fig: M-R RR}
\end{figure*}
The self-gravitating compact objects formed with Bose-Einstein condensates are known as BEC stars~\cite{Jones_2001,Chavanis:2011cz}. Properties of these stars have been investigated within the framework of GR. BEC stars are also proposed to be possible candidates for elusive dark matter~\cite{Boehmer:2007um}.  
BEC stars at finite chemical potential and temperature are also considered~\cite{Bhatt:2009gv,Latifah:2014ima,Aswathi:2023zzn}. 
In order to study the global properties of zero and finite temperature BEC stars in modified gravities, we consider the following popular descriptions of BEC matter. A relativistic BEC equation of state (EoS), known as the Colpi-Shapiro-Wasserman (CSW), by considering scalar field theory with $\phi^4$ interactions is given as~\cite{Colpi:1986ye}
\begin{equation}
\label{CSW}
p(\rho)=\frac{1}{36K^{'}}\left\lbrack \left (1+12K^{'}\rho \right )^{1/2}-1\right\rbrack^2,
\end{equation}
with $K^{'}= \lambda \hbar^3/(4m^4)$. Here, $\rho$ represents the mass density and $m$ is the mass of condensate particle. $\lambda$ is a dimensionless quantity defined by 
$\lambda = (9.523 \times{8\pi}) \left( \frac{a}{1\, {\rm fm}}\right)\left(\frac{m}{2 m_n}\right)$, where $m_n$ is the mass of a nucleon and $a$ denotes the scattering length of bosons in the system. 
Another widely used BEC EoS based on the Gross-Pitaevskii equation is given as~\cite{Chavanis:2011cz} 
\begin{equation}\label{Ch}
p(\rho)=K\rho^2=\frac{g\rho^2}{2m^2},
\end{equation}
where $g=4\pi a\hbar^2/m$ denotes the interaction strength. Further, we study the BEC stars with finite temperature, using a recently developed EoS~\cite{Gruber:2014}
\begin{align} \label{eos}
        p(\rho,T) =&\frac{g\rho^2}{2m^2} + \frac{2g\rho}{m\lambda_{th}^3}\,\zeta_{3/2}\left[e^{- g\rho/(mk_BT)}\right]  \\
        &+\frac{2k_BT}{ \lambda_{th}^3} \zeta_{5/2} \left[e^{-g\rho/(mk_BT)}\right] -\frac{2k_BT}{\lambda_{th}^3}\,\zeta_{5/2}\left[ 1\right]. \nonumber
\end{align}
Here, $T$ represents the system temperature and $k_B$ is the Boltzmann constant. The thermal de Broglie wavelength is given by $\lambda_{th}=\sqrt{2\pi\hbar^2/mk_BT}$ and 
 $\zeta_\nu [z]$ represents the polylogarithmic function of order $\nu$ defined as $\zeta_{\nu} [z]=\sum_{n=1}^\infty z^n/n^\nu$. 
 The first term of the equation stated above represents the zero temperature result given by Eq.~(\ref{Ch});  
 while rest of the terms arise because of the presence of temperature. 
The above mentioned three BEC prescriptions have been used to study  static~\cite{Boehmer:2007um,Chavanis:2011cz,Gruber:2014} and rotating~\cite{Mukherjee:2014kqa,
Chavanis:2011cz,Aswathi:2023zzn} BEC stars based on Newtonian gravity and GR. We take the mass of condensate particle as $m=2m_n$, with the nucleon mass $m_n=1.675\times 10^{-24}$ g; this accounts for the possibility that the two nucleons forming an equivalent Cooper pair and existing as a boson. We choose the scattering length to be $a=1$ fm and correspondingly, the interaction strength takes the value $g=4.17 \times 10^{-43}$ g cm$^5$/s$^2$ in our analysis. 
 Using these BEC matter prescriptions in Eqs.~\eqref{rhotilde} and~\eqref{ptilde}, we obtain the effective pressure and density (for  given values of $\kappa$ and $\Sigma$) to solve the stellar structure equations for a BEC star. 
 \par 
Next, by numerical methods, we proceed to obtain the stellar configurations for zero and finite temperature BEC stars in the combined RR gravity. 
We solve the coupled differential equations: Eqs.~(\ref{eq:TOVPR}) and~(\ref{eq:TOVMR}) by choosing different values for the Rastall parameter $\kappa$ and the Rainbow function $\Sigma$ for the BEC EoSs i.e., Eqs.~(\ref{CSW}), \eqref{Ch} and (\ref{eos}). As noted before, for the $\Sigma=\kappa=1$ case, we recover the TOV solutions of GR. 
\par 
In order to study the behaviour of stellar profiles with the parameters $\kappa$ and $\Sigma$, we first consider BEC stars within the Rastall and Rainbow gravities by choosing the values of these parameters accordingly. 
In Fig.~\ref{MR-Rastall}, we consider the zero temperature BEC star using the relevant EoSs, i.e., Eqs.~(\ref{CSW}) and (\ref{Ch}), denoted as - CSW and $K\rho^2$ respectively, in the Rastall theory (by keeping the Rainbow function $\Sigma=1$). The mass-radius curves are studied for different values of the Rastall parameter $\kappa=0.995,\,1.000,\,1.005,\,1.050$. As seen from the figure, $K\rho^2$ EoS results in higher maximum masses compared to CSW; although much difference in the corresponding radii is not seen. The maximum mass of the BEC star within GR ($\kappa=1$) is obtained to be 0.44\,$M_\odot$ (0.71\,$M_\odot$) for the CSW ($K\rho^2$) EoS and the corresponding radius is 4.05\,km (3.98\,km). 
When $\kappa$ is varied from unity, no appreciable change can be noticed in the maximum mass for either the EoSs, whereas the corresponding radius shows minor increment with $\kappa$. 
The deviation from GR is noticeable even for small increments in $\kappa$ ($1$ to $1.005$), especially for stars with $\rho_c < 7 \times10^{15} \,$g cm$^{-3}$.  
At low densities, the $K\rho^2$ and CSW EoSs result in almost equal mass and radius. Further, when the Rastall parameter $\kappa$ is increased to $1.05$ and beyond, no stable BEC stellar solutions are possible (for either the EoSs), as seen in the case of neutron stars \cite{Mota:2019}. 
\par
Fig.~\ref{MR-Rainbow} represents a similar analysis within the Rainbow formalism by keeping $\kappa=1$. Here too, as observed in the Rastall theory, the CSW EoS results in lower mass-radius curves compared to $K\rho^2$. One can see that, a small increment in the Rainbow function $\Sigma$ results in an appreciable enhancement of the mass-radius curves, along with values of maximum mass and corresponding radius.  
The deviation from GR stellar equilibria ($\Sigma =1$) is visible distinctly for all the central densities, for both the EoSs.  
It can be seen that the behaviour of mass-radius curves are (almost) qualitatively  similar for both the EoSs, except that $K\rho^2$ results in higher mass and radius for a given value of central density. 
The maximum mass of $K\rho^2$ (CSW) for $\Sigma = 0.95,\,1.05$ and $1.2$  are $0.68 M_\odot$ ($0.42 M_\odot$), $0.75 M_\odot$ ($0.46 M_\odot$) and $0.86 M_\odot$ ($0.52 M_\odot $) respectively. 
Here, we also note that if either $\kappa$ or $\Sigma$ is chosen to be less than unity, the resultant solution shows an overall reduction in the mass and corresponding radius. Furthermore, if $\kappa$ is less than unity, we find that certain values of effective pressure may become negative and we avoid such scenarios in our analysis~\cite{Mota:2019}. 
\par
Now, we turn our attention towards the Rastall-Rainbow theory. In doing so, we include the effect of temperature too into the analysis by considering the finite temperature EoS described in Eq.~(\ref{eos}); which becomes $K\rho^2$ in the zero temperature limit. 
 \begin{figure}[t]
    \centering    \includegraphics[width =\linewidth]{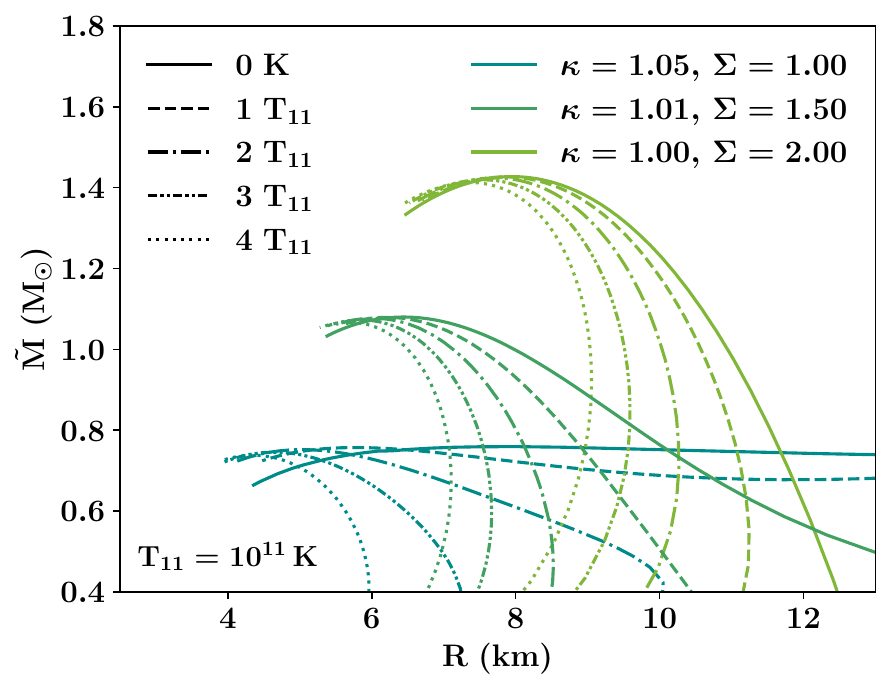}
    \caption{Effect of temperature on BEC stellar configurations within Rastall, Rainbow and Rastall-Rainbow gravities. 
    }
    \label{fig:RR-T}
\end{figure}
In Fig.~\ref{fig:RR-T}, we plot the mass-radius profiles of BEC stars at relevant different temperatures 
$(1T_{11}-4T_{11})$, where $T_{11}=10^{11}\,$ K for Rastall, Rainbow and combined RR gravities. 
Firstly, we study the effect of temperature on the stellar configuration of Rastall BEC star by fixing the Rainbow function $\Sigma =1$ and for a fixed value of Rastall parameter, $\kappa=1.05$. As noted earlier, at $0\,$K, stable BEC stellar solutions are not obtained in the Rastall theory when $\kappa$ increases to $1.05$.  
However, we infer from Fig.~\ref{fig:RR-T} that, for the same value of $\kappa\,(1.05)$, it is possible to obtain stable stellar configurations in the Rastall gravity with the increase in temperature. With $\kappa = 1.05$, we obtain a BEC star with maximum mass $\tilde{M}=0.75 M_\odot$ having radius $R=5$ km at $T=2T_{11}$. Similarly, we study the impact of temperature on BEC stellar profiles within Rainbow gravity by fixing $\kappa =1$ and choosing $\Sigma=2$. We observe only a marginal change in the maximum mass and radius of the star with the increase in temperature. Further, the stellar equilibria with the same mass are observed to have reduced radii with the increase in temperature. Now, in order to study the impact of temperature within RR theory, we choose $\kappa=1.01$ and $\Sigma=1.5$. The maximum mass of the star for different temperatures is obtained to be $\approx 1.07 M_\odot$ and the corresponding radius is $\approx 6$ km. We find that, for the chosen values of $\Sigma$ and $\kappa$, we do not obtain unstable configurations for all the temperatures considered. 
Moreover, it must be noted that, the presence of temperature has only minimal effect on the maximum mass and radius of the star in all the three theories of gravity considered here. This observation is inline with the result obtained for the case of static and rotating BEC stars at finite temperature within Einstein's GR~\cite{Aswathi:2023zzn}.
\begin{figure*}
  \centering
  \subfigure
  {\includegraphics[scale=0.8]{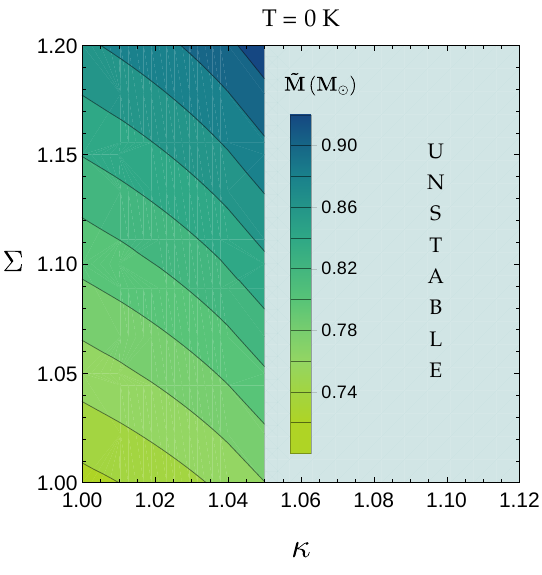}}\quad
  \subfigure{\includegraphics[scale=0.8]{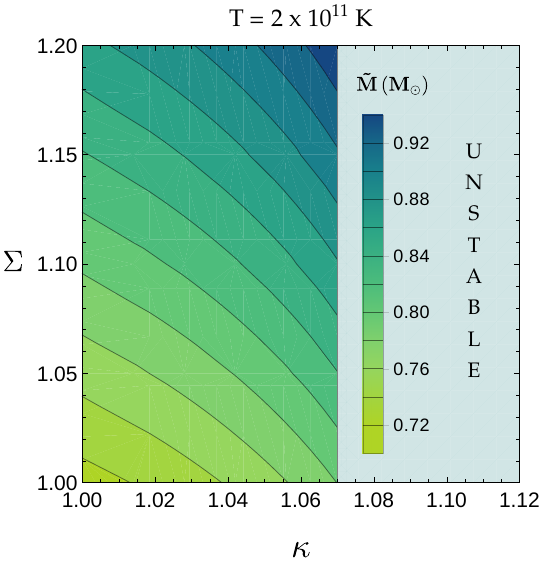} 
  } 
  \caption{Maximum mass of the BEC star $\tilde M$ as a function of Rastall parameter $\kappa$ and Rainbow function $\Sigma$.}
  \label{fig:RR-contour}
\end{figure*}
\par 
In Fig.~\ref{fig:RR-contour}, we plot the constant maximum mass contours of BEC stars within the Rastall-Rainbow theory at zero and finite temperatures ($2T_{11}$). As observed in the Rastall formalism, at zero temperature, stable BEC star configurations are not obtained in the RR gravity with $\kappa = 1.05$ and beyond, for any value of $\Sigma$. However, including temperature in our analysis through Eq.~\eqref{eos} is found to extend the parameter space by allowing more range of values for $\kappa$.
At $T=2\times 10^{11}$ K, we obtain stable BEC stellar configurations upto $\kappa=1.07$, for all values of $\Sigma$ considered. 
Moreover, it must be noted that the effect of Rainbow function is to vary the maximum mass values of the stellar configurations. With an increment in $\Sigma$, 
configurations with larger maximum mass values can be achieved;
while, the Rastall parameter affects the stability of the solutions. 
\par
\begin{figure}
    \centering  
\includegraphics[scale=0.46]{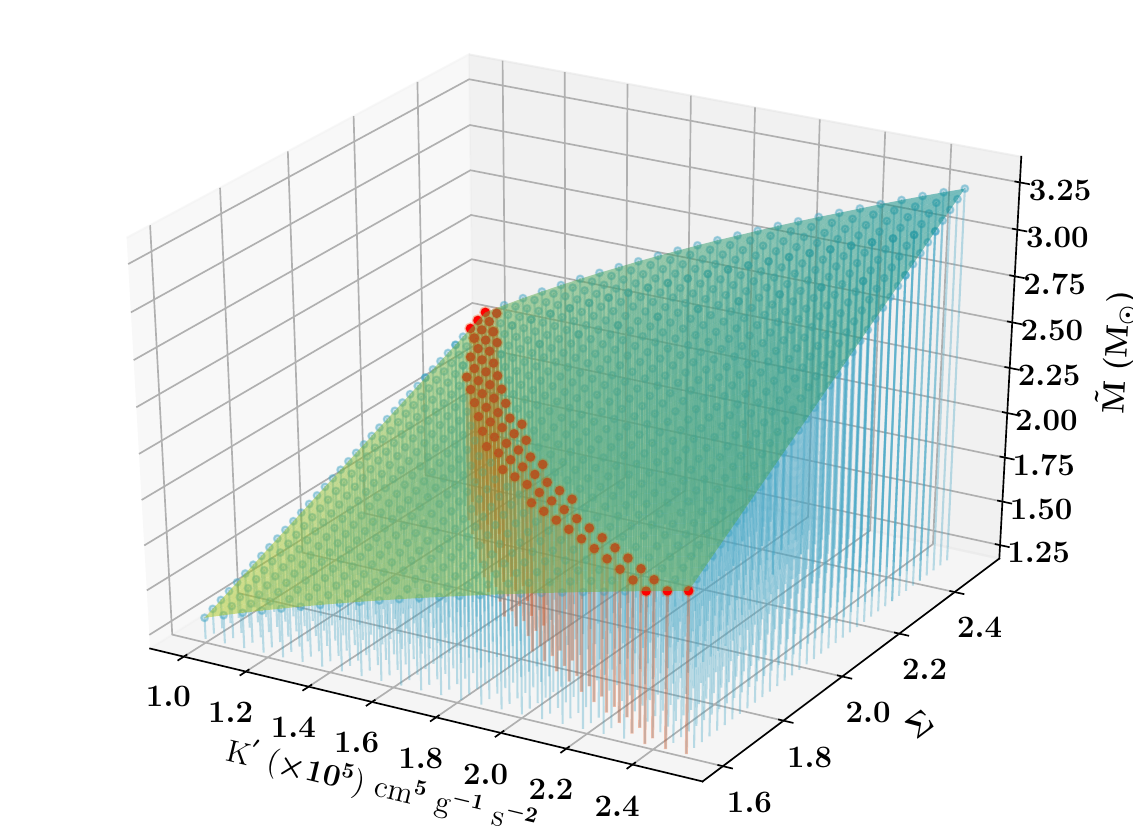}
    \caption{BEC stars obeying CSW EoS within RR gravity by varying the EoS parameter $K^{'}$ and Rainbow function $\Sigma$ for a fixed value of Rastall parameter $\kappa = 0.95$. The maximum mass and radius values which are in agreement with NICER observations ($2.14_{-0.18}^{+0.20} M_{\odot}$ 
 and $13.7_{-1.5}^{+2.6}$ km) of PSR J0740+6620~\cite{NANOGrav:2019jur,Miller:2021qha} are shown in red. }
    \label{fig:RR-csw}
\end{figure}

We now compare our results obtained within the Rastall-Rainbow theory for the three EoSs with the observational predictions. Reference~\cite{NANOGrav:2019jur} reports the existence of the millisecond pulsar MSP J0740+6620 having mass $2.14_{-0.09}^{+0.10} M_{\odot}$ (upto $68.3\%$ credible) and $2.14_{-0.18}^{+0.20} M_{\odot}$ ($95.4\%$ credible) as found by the Neutron star Interior Composition Explorer (NICER). The NICER data also provides the equitorial radius of PSR J0740+6620 to be $13.7_{-1.5}^{+2.6}$ km ($68\%$ credible)~\cite{Miller:2021qha}. 
As mentioned earlier, a BEC star obeying the CSW EoS within Einstein's GR is found to have a maximum mass $\tilde{M} = 0.44 M_\odot$ and corresponding radius $R = 4.05$ km keeping $a=1$ fm and $m=2m_n$ ($K^{'} = \lambda \hbar^3/(4m^4)=1.859\times10^{4} $cm$^5\,$g$^{-1}\,$s$^{-2}$). Such a BEC stellar configuration is inconsistent with the above data. By keeping $K^{'}\geq 3.475\times10^{5} $cm$^5\,$g$^{-1}\,$s$^{-2}$, within GR one gets BEC star with maximum mass in agreement with the above observation; however, the corresponding radius is found to exceed the generally allowed range of values~\cite{Mukherjee:2014kqa}. Therefore, all configurations of the BEC star obeying CSW EoS are ruled out within GR due to the observational constraints on the radius values. Now, coming to the RR gravity, one can obtain BEC stellar configurations using the CSW EoS, with gravitational masses and radii consistent with observational predictions. We obtain stars with maximum masses $\approx 2 M_\odot$ and the corresponding radii in the range $15.0-16.3$ km by varying the values of free parameters of the theory $(\kappa, \Sigma)$ and $K^{'}$. 
We plot such BEC star 
configurations for a fixed value of $\kappa = 0.95$ in Fig~\ref{fig:RR-csw}. 
\par
Coming to the $K\rho^2$ EoS, by varying $\Sigma$ from $2.7$ to $3.0$ with $\kappa=1$, we get BEC stars with
maximum masses lying in the range $1.92 M_\odot$ - $2.14M_\odot$; which are in agreement with the observational constraints \cite{Antoniadis:2013pzd,Demorest:2010bx,Romani:2022jhd,NANOGrav:2019jur}.
 For the finite temperature ($2T_{11}$) case, within General Relativity, we obtain BEC stellar configurations having maximum masses in the range 2.0$M_{\odot}$ - 2.2$M_{\odot}$ which are consistent with the observations, only when we keep the interaction strength to be large enough i.e., $8g-10g$. However, in RR gravity, with a lower interaction strength of $5g$, we can obtain stable BEC star configurations with maximum masses in the range 1.9$M_{\odot}$ - 2.2$M_{\odot}$ (with corresponding radii in the range $9.9-13.1$ km) for the Rastall parameter values 1$\leq\kappa\leq$1.07 and by varying $\Sigma$ from $1.2$ to $1.3$. 
 %

\par 
In conclusion, we have studied the global properties of static BEC stars within the Rastall, Rainbow and combined Rastall-Rainbow (RR) theories of gravity. We have employed three different EoSs and considered both zero and finite temperature BEC stars in our analysis. The mass-radius profiles of the BEC star in General Relativity (GR), Rastall, Rainbow and RR theories of gravity are obtained by numerically solving the modified stellar structure equations, by choosing Rastall parameter $\kappa$ and Rainbow function $\Sigma$ accordingly. 
We find that, within these modified theories, the BEC stellar equilibria show difference in masses and radii with minimal changes in the free parameters, implying deviation from the GR case. 
We observe that, increasing the value of $\kappa$ beyond certain limit results in unstable solutions irrespective of the value chosen for $\Sigma$. 
However, it can be seen that the maximum mass and corresponding radius of stable BEC stars in Rastall gravity are not sensitive to the 
changes in $\kappa$, whereas in the Rainbow formalism, the parameter $\Sigma$ varies them appreciably. 
The inclusion of temperature is observed to have minimal effect on the maximum mass and radius of the stable stellar configurations in all three modified theories considered. 
Also, we have found that the presence of temperature expands the parameter space in the RR theory by allowing more range of values for $\kappa$ that results in stable stellar equilibria. 
\par 
We report that, all the three EoSs considered here allows to have BEC stars in the RR gravity with maximum masses and radii in good agreement with the recent pulsar observational data. 
In particular, the Colpi-Shapiro-Wasserman EoS allows the existence of BEC stars consistent with the observational constraints within RR theory, unlike the case of GR. 
Further, we find that, in contrast to the predictions
 of GR, BEC stars having masses consistent with observational data can be obtained in the RR theory for smaller interaction strength of bosons, by careful selection of the 
 parameters $\kappa$ and $\Sigma$. 
\par
L. J. N. acknowledges the Department of Science and Technology, Government of India for the
INSPIRE Fellowship.
\bibliography{RR-BEC}
\end{document}